\documentclass[preprintnumbers,article,amsmath,amssymb,floatfix,10pt,prd,onecolumn,
superscriptaddress,nofootinbib]{revtex4}
\usepackage[colorlinks=true, pdfstartview=FitV, linkcolor=blue, citecolor=red, urlcolor=magenta]{hyperref}
\usepackage{bbm}
\usepackage{amsfonts}
\usepackage{mathrsfs}
\usepackage{latexsym}
\usepackage{epsfig}
\usepackage{epstopdf}
\usepackage{epstopdf}
\usepackage{graphicx}
\usepackage{amssymb}
\usepackage{amsmath}
\usepackage{dcolumn}
\usepackage{bm}
\usepackage{color}
\usepackage{comment}
\usepackage{xcolor}
\usepackage{graphicx}
\usepackage{dcolumn}
\usepackage{bm}
\usepackage{color}
\usepackage{enumitem}
\usepackage{amsmath}
\usepackage{amssymb}

\usepackage[T1]{fontenc}
\usepackage[utf8]{inputenc}   
\usepackage{lmodern}
\usepackage{cmap}

\begin{document}
\title{\bf Impact of deformation parameters on the greybody factor and Joule-Thomson expansions of Schwarzschild-AdS black hole}

\author{Faisal Javed}
\email{faisaljaved.math@gmail.com}\affiliation{School of Mathematical and Statistics, Zaozhuang University, Zaozhuang, 277160, China}
\affiliation{Research Center of Astrophysics and Cosmology, Khazar University, Baku, AZ1096, 41 Mehseti Street, Azerbaijan}
\affiliation{Department of Mathematics, School of Science, University of Management and Technology, Lahore, 54000, Pakistan}

\author{A. Eid}
\email{amaid@imamu.edu.sa} \affiliation{Department of Physics, College of Science, Imam Mohammad Ibn Saud Islamic University (IMSIU), Riyadh, Kingdom of Saudi Arabia}

\author{Arfa Waseem}
\email{arfa.waseem@gcwus.edu.pk} \affiliation{Department of
Mathematics, Government College Women University, Sialkot, Pakistan}

\author{Jiafeng Guo}
\email{jiafeneg_guo@163.com}\affiliation{Department of Mathematical Sciences, Zibo Normal College, Zibo 255130, China.}

\author{Abdelmalek Bouzenada}
\email{abdelmalekbouzenada@gmail.com} \affiliation{Laboratory of Theoretical and Applied Physics, Echahid Cheikh Larbi Tebessi University 12001, Algeria}

\author{Sulaman Shaukat}
\email{sulamanshaukat444@gmail.com}\affiliation{Department of Mathematics, University of the Punjab,
Quaid-e-Azam Campus, Lahore-54590, Pakistan.}

\author{Farruh~Atamurotov}
\email{atamurotov@yahoo.com}
\affiliation{University of Tashkent for Applied Sciences, Str.Gavhar 1, Tashkent 100149, Uzbekistan}
\affiliation{Urgench State University, Kh. Alimdjan str. 14, Urgench 220100, Uzbekistan}

\author{N. Mustapha}
\email{nmmustapha@imamu.edu.sa} \affiliation{Department of Physics, College of Science, Imam Mohammad Ibn Saud Islamic University (IMSIU), Riyadh, Kingdom of Saudi Arabia}

\begin{abstract}
This paper investigates the greybody factor and Joule-Thomson expansion of deformed Schwarzschild-AdS black hole models. We constructed the Klein-Gordon equation into a Schrödinger-like wave equation using tortoise coordinates and derived the exact effective potential. Also, we analyzed its behavior under varying physical parameters, including deformation parameters, mass and angular momentum. In this case, our results illustrate that increasing the deformation parameter enhances the effective potential, significantly impacting the absorption and scattering of massless scalar fields. It is noted that the Schwarzschild black hole effective potential is greatly increased when the deformation parameter is present. Near the event horizon, we found that waves behave differently, with higher-frequency waves more easily penetrating potential obstacles and lower-frequency waves bouncing off them. Furthermore, we investigated the thermodynamic characteristics of black holes using the Joule-Thomson expansion, detecting isenthalpic curves that indicate heating and cooling zones controlled by pressure and temperature changes. In this context, we tested the influence of deformation black hole parameters; the isenthalpic curves are enhanced by increasing the deformation parameters and the black hole mass.
\end{abstract}
\maketitle

\date{\today}


\section{Introduction}\label{S1}

In the study of gravitational influence measurement, black holes (BH)s have fascinated physicists for decades because they offer an exceptional arena where the laws of fundamental physics converge in their most extreme form, providing a deeper explanation and revealing the structure of spacetime and matter interaction. Far from being mere astrophysical measurements, they now stand as central objects in the exploration of thermodynamic properties under conditions unattainable elsewhere in the universe, serving as laboratories that illustrate the limits of general relativity (GR), quantum mechanics and classical thermodynamic laws. Another important result in BH models is that the study of BH thermodynamics has evolved into one of the most compelling interdisciplinary fields of modern theoretical physics, where phenomena such as entropy, temperature and phase transitions are reinterpreted in a gravitational framework, opening paths toward resolving one of the greatest scientific challenges, the formulation of a consistent theory of quantum gravity (QG). Also, at the heart of this framework lies the event horizon, a fundamental boundary beyond which no form of matter or radiation can escape, which not only defines the causal structure of the BH but also illustrates crucial thermodynamic information. In this case, Bekenstein \cite{Bk} first recognized that BHs must possess entropy proportional to the area of their event horizon, thereby overturning the earlier view of BHs as mere cosmic sinks of mass and energy and instead promoting them to thermodynamic entities with a well-defined informational content. Also, this result shows and illustrates the foundation for Hawking’s subsequent discovery of BH radiation, ultimately reshaping our perspective of BHs as dynamic systems governed by principles analogous to those of ordinary thermodynamics, with the horizon area serving as a measure of disorder and information storage. In this context, the thermodynamic interpretation of BHs enhances our understanding of their internal dynamics as well as stability and explains the relationship between gravitation, entropy and quantum phenomena, thereby bridging the classical and quantum domains in the ongoing search for a unified physical theory. 

In 1974, Hawking tested the work of Bekenstein by illustrating that BH properties, long considered to be perfect absorbers from which no information or energy could escape, actually emit thermal radiation as a consequence of quantum field effects in curved spacetime (testing the influence of curved spacetime on particle interaction), a phenomenon now universally known as Hawking radiation. This result arose from a careful analysis of quantum vacuum fluctuations near the event horizon, where particle-antiparticle pairs are continuously created by the uncertainty principle. In this context, one member of the pair can fall inside the event horizon while the other escapes to infinity, producing a net outward flux of radiation that slowly drains the BH of its mass and eventually leads to complete evaporation \cite{Haw}. Also, the discovery had immense implications, as it not only bridged the gap between GR and QG but also overturned the long-standing belief that nothing, not even light, could ever cross outward from the event horizon. Another important result in BH thermodynamics measurement, Hawking's work revealed that BHs are characterized by a well-defined temperature, inversely proportional to their mass, and thus behave as thermodynamic systems with finite temperature and entropy, thereby embedding them within the broader framework of statistical mechanics \cite{Wad}. Also, this revolutionary insight confirmed and strengthened Bekenstein’s earlier proposal that the entropy of a BH is proportional to the surface area of its event horizon, and together, Bekenstein and Hawking defined and explained the foundations of BH thermodynamics. In this case, their framework established the four fundamental laws that mirror the classical laws of thermodynamics: the zeroth law, which identifies the constancy of surface gravity (analogous to temperature) across the event horizon; the first law, which relates changes in mass, area and angular momentum to energy conservation; the second law, which dictates that the total entropy, defined through the horizon area, can never decrease and the third law, which asserts that it is impossible to reduce the surface gravity to zero by finite processes. Also, the generalized second law remains one of the most significant, explaining that the entropy of a BH plus the entropy of its surroundings cannot diminish, preserving consistency with the broader thermodynamic principle that entropy in a closed system is non-decreasing \cite{Georg}. In this case, this unification of concepts gives new perspectives on the ultimate fate of BHs and illustrates the nature of QG, presenting that BHs are thermodynamic objects whose temperature, entropy, and energy conservation are dictated by deep principles that intertwine gravity, thermodynamics and quantum mechanics at the most fundamental level. 

Also, a significant body of work has revealed that BHs can be consistently interpreted within a thermodynamic framework, thereby drawing a compelling analogy between gravitational dynamics and conventional thermal systems \cite{P1,P2,P3}. In this case, this perspective has its roots in attempts to decode the implications of Einstein’s field equations, where quantities traditionally associated with thermodynamics, such as temperature, specific heat, and entropy, find natural counterparts in BH physics, illustrating new information about BH structure and QG \cite{Q1,Q2,Q3}. Also, the relation between generalized Born-Infeld (BI) electrodynamics and string theory has been tested in \cite{md1}, pointing toward deep theoretical connections beyond classical GR. Within the extended phase space formalism, the role of the cosmological constant is reconceptualized as a dynamical pressure term, leading to a reinterpretation of the BH mass not as internal energy but rather as the enthalpy of the gravitational system. In the anti-de Sitter (AdS) setting, this framework naturally introduces a conjugate variable corresponding to the thermodynamic volume of the BH, thereby enriching the analogy between gravitational and thermodynamic variables \cite{A1}. Also, some studies have examined AdS BH models within higher-order curvature modifications, such as Lovelock gravity \cite{A2}, where explicit formulations for the mass and free energy have been developed, shedding light on the intricate thermodynamic behavior of these spacetimes \cite{A3}. Also, investigations into dilaton-AdS BHs \cite{A4} showed that the physical mass of the system is most appropriately interpreted as enthalpy, consolidating the viewpoint that the extended thermodynamic phase space illustrates a natural and unifying description. These studies have substantially expanded our comprehension of the thermal properties and phase transitions and have explained theoretical models of AdS BHs, reinforcing their central role as laboratories for probing fundamental aspects of gravitational physics in extended thermodynamic settings.

Extended phase space thermodynamics provides a systematic
approach to studying the thermodynamic properties of black
holes in AdS spacetime. In this formulation, the cosmological
constant $\Lambda$ is allowed to vary and is identified with
the thermodynamic pressure, while its
conjugate variable is the thermodynamic volume $V$. This
description permits an analysis of BH phase transitions
and critical phenomena within a framework analogous to
conventional thermodynamics. In particular, certain AdS
BH solutions exhibit phase behavior similar to that
of a van der Waals fluid, including small/large BH
transitions and critical points \cite{M1,M2,F1,F2,F3}. These
similarities establish a connection between BH
thermodynamics and the thermodynamics of interacting fluids.
The AdS/CFT correspondence provides a further interpretation
of these phenomena by relating the thermal properties of AdS
BHs to those of the dual conformal field theory
\cite{AdS}.

The emission of Hawking radiation is another important aspect
of BH thermodynamics. For a neutral, massless scalar
field, the thermal occupation number is determined by the
Hawking temperature $T_H$ through the Bose-Einstein
distribution \cite{Haw}.
The thermal character of Hawking radiation is
central to the information-loss problem, which concerns
whether BH formation and evaporation preserve
quantum information. The radiation spectrum is modified as the field propagates
through the curved spacetime outside the event horizon.
The effective potential partially transmits the outgoing
radiation and reflects the remaining part towards the black
hole. This scattering is characterized by the greybody
factor (GBF), $|A_{l,m}(w)|^2$, which gives the transmission
probability for each angular mode. In the corresponding
absorption problem, it represents the fraction of incident
flux absorbed by the BH and is related to the
absorption cross-section \cite{15'u}-\cite{18u}. 
Thus, the frequency-dependent GBF accounts for the departure
of the emitted spectrum from an ideal thermal distribution
due to scattering by the exterior geometry. In AdS spacetime,
the interpretation of this flux also depends on the boundary
conditions imposed at the conformal boundary.  Extended phase space thermodynamics provides a framework for
studying the thermodynamic properties and phase transitions of
AdS BHs. In this approach, the cosmological constant is
allowed to vary and is identified with the thermodynamic pressure,
$P=-\Lambda/(8\pi)$, whose conjugate variable is the thermodynamic
volume $V$. The BH mass is consequently interpreted as
enthalpy. For suitable BH solutions, the resulting phase
structure exhibits features analogous to those of a van der Waals
fluid, including small/large BH phase transitions and
critical behavior \cite{M1,M2,F1,F2,F3}. The AdS/CFT correspondence
provides a related interpretation of AdS BH thermodynamics
in terms of thermal states and phase transitions in the dual
conformal field theory \cite{AdS}.

The Joule-Thomson (JT) effect is a principle of classical thermodynamics that elucidates the behavior of a gas transitioning from a high-pressure zone to a low-pressure region at a constant velocity. Also, this phenomenon was first studied in AdS BH models \cite{J1}, but has since been thoroughly analyzed across various BH model measurements \cite{J4, J6, J7, J8, pd10b}. In this case, a comprehensive examination revealed that heating and cooling areas may be seen on a temperature-pressure phase diagram. Also, the positions of these zones are established by the intersection of the enthalpy and inversion curves; specifically, a heating area is linked to a negative JT factor, while a cooling region is related to a positive coefficient. In this context, a crucial element in the JT effect is the ratio of the minimal inversion temperature to the critical temperature. Also, the ratio for Reissner-Nordström-AdS (RN-AdS) BH models is $1/2$. Another important result is that previous studies have evaluated this ratio for other notable BHs \cite{J2, J3, J5}.

The analysis of BHs is regarded as the most sophisticated domain of inquiry in the scientific community, owing to the distinctive attributes shown by the thermodynamic events in both classical and quantum gravitational frameworks. In this case, classical theories assume that BHs exert a significant gravitational pull that impedes the flow of particles and radiation over the event horizon. Also, this process ultimately leads to the gradual destruction of BHs by the emission of Hawking radiation \cite{q1}. As a quantum effect, Hawking radiation causes BHs to lose mass over time until they eventually evaporate completely. The relationship between BH thermodynamics and classical thermodynamics is illustrated by differences in Hawking temperature and entropy between various kinds of BHs. The connection among the mass of BHs, temperature, energy, and surface gravity is further explained via this link. Also, showing BHs thermal characteristics and connection to event horizons requires knowledge of their entropy. Also, to investigate thermal fluctuations, which are crucial to the difficulties in achieving thermal equilibrium between BHs and thermal radiation, Bekenstein \cite{q3} proposed logarithmic modifications in the entropy-area relationship. In the RN framework, Kanzi and Alipour \cite{jc1} investigated the thermal properties, including entropy, heat capacity, and energy, in relation to the dynamics of Dirac particles. In this case, in the holographic framework, the RN spacetime encodes four information bits for Dirac (spin-1/2) and  Klein-Gordon (KG) (spin-0) particles \cite{jc1}. Also, further studies showed that quantum oscillations affect BH geometry, while thermal fluctuations from statistical disturbances reveal properties of celestial bodies. Also, a study of Kerr-Ads BH was tested by Pourhassan and Faizal (for more information and detail, see Ref. \cite{q3a}). This study employed a logarithmic correction to enhance BH entropy and investigated the effects of thermal variations on thermodynamic potentials. 

Across different higher-dimensional BH models, it was observed that the inclusion of higher-order correction terms plays a decisive role in shaping the outcomes \cite{q3b, q3c}. Pourhassan et al. \cite{q3d} showed how logarithmic corrections alter essential thermodynamic quantities such as entropy, heat capacity, and the Hawking temperature for the modified Hayward BHs. In another study, Ali et al. \cite{gt1} tested the extended phase space and thermodynamic behavior of a magnetically charged Euler-Heisenberg BH in AdS space coupled minimally with a scalar field. Mola et al. \cite{gt2} focused on the influence of strong gravitational lensing, quasi-periodic oscillations, and constraints derived from $\mathcal{EHT}$ observations on quantum-corrected charged BHs. Abbas et al. \cite{gt3} considered extended phase space thermodynamics in the context of \textsl{AdS} hairy BHs within the gravitational decoupling framework. Also, Ditta et al. \cite{gt4} analyzed the thermodynamic profile of the Simpson-Visser Minkowski core BH solution in relation to Verlinde’s emergent gravity perspective. In this case, the influence of a surrounding cloud of strings on different properties of BHs within various gravitational theories has also been extensively discussed by several researchers \cite{f4,f5,f6,f7,f8,f9}.

Jawad and Shahzad \cite{q3e} explored non-minimal regular BHs by incorporating the cosmological constant, showing that ordinary BHs remain stable when this constant is increased. In another contribution, Zhang \cite{q3f} tested how first-order entropy corrections together with thermodynamic quantities influence Kerr–Newman–\textsl{AdS} and \textsl{RN} BHs. Also, the tunneling of vector particles and the associated Hawking temperature within three-dimensional Gödel BHs were addressed in \cite{q3g}. The issue of charged BH stability under thermal fluctuation effects was considered in \cite{q3h}. Also, the thermodynamic behavior of rotating charged \$(2+1)\$-dimensional BHs was studied in \cite{c1}, offering additional insights into their physical properties.

Ditta et al. \cite{pd1} investigated the thermodynamic behavior of symmergent BHs. The connection between thermal oscillatory effects and Schwarzschild BH within scalar–tensor gravity was discussed in \cite{pd2}. The role of nonlinear electrodynamics in the JT expansion process of charged BHs incorporating higher-order contributions was studied in \cite{pd3}. Also, the work in \cite{pd4} explored how thermal fluctuations modify the thermodynamics of charged $BTZ$ BHs. A comprehensive treatment of free energy in the framework of string theory is given in \cite{pd5}. Moreover, the influence of thermal corrections on quantum-improved charged BHs is thoroughly examined in \cite{f25}. In this context, additional research considers the thermodynamics of BHs, the \textsl{GBF} approach, and fluctuation effects across various modified gravity models \cite{pd6}–\cite{pd9}.

Using the method of gravitational decoupling, Khosravipoor and Farhoudi \cite{BH} have obtained the solution of the deformed Schwarzschild-AdS BH, where they have considered an additional gravitational source, which adheres to the energy principle of weak gravity. In particular, they chose energy density as a monotonic function that conforms to the available requirements. This method determines the influence of geometric distortions on the geometry itself through the positive deformation parameter \cite{BH}, which has a certain upper limit in the scenario of the event horizon. By making the deformation parameter take on a value of zero, the thermodynamic qualities of the BH revert to those that were recorded in earlier studies. {Although greybody factors and JT expansion describe different physical phenomena, both provide complementary information about the properties of a BH. The greybody factor characterizes the scattering and absorption of scalar waves as they propagate through the effective potential generated by the BH geometry, whereas the JT expansion describes the thermodynamic response of the BH during an isenthalpic pressure variation in the extended phase space. In the present deformed Schwarzschild-AdS model, both analyses are based on the same metric function, which incorporates the deformation parameters $\alpha$ and $\beta$. These parameters modify the spacetime geometry and can consequently influence both the effective potential governing wave propagation and the thermodynamic quantities determining the isenthalpic and inversion curves. Investigating these two aspects together therefore provides a broader assessment of the physical consequences of spacetime deformation, connecting the wave-scattering properties of the BH with its thermodynamic response without assuming a direct mathematical coupling between them.}

This paper aims to observe the \textsl{GBF} and the JT expansion of the deformed AdS Schwarzschild BH. In section (\ref{S2}), the deformed AdS-Schwarzschild geometry shall be investigated with an emphasis on the KG equation and the holographic potential of a massless scalar field. The equation of motion (EoM) in radial is built, which is then followed by an analysis of the characteristics of the effective potential. Section (\ref{S3}) centers on the \textsl{GBF}, furnishing the analytical solution to the near-horizon (NH) and asymptotic regions by solving the radial equation. In section (\ref{S4}), a comparison of the two regions' data is given, and a computation of the emission and absorption rates of a massless scalar field is performed. In section (\ref{S5}), the JT expansion is considered, and the importance of the deformation parameter on inversion and isenthalpic curves is stated. In this context, the last part presents the conclusion about our results with an explanation of the influence of the BH parameters.

\section{Klein-Gordon equation and effective potential of Deformed Schwarzschild-AdS Black Hole models}\label{S2}

In this part, we begin with a static, spherically symmetric line element as the foundation for constructing the modified BH configuration, expressed as
\begin{equation} \label{final_line_element}
ds^{2}=-f(r)dt^{2}+\dfrac{1}{f(r)}dr^{2}+d\mathcal{S}_{2\mathcal{D}}^{2}.
\end{equation}
and the angular sector $d\mathcal{S}_{2\mathcal{D}}^{2}$ of the metric is given by
\begin{equation}
d\mathcal{S}_{2\mathcal{D}}^{2}=r^{2}\left(d\theta^{2}+\sin^{2}\theta\,d\phi^{2}\right),    \end{equation}
where the radial function defining the deformed-AdS BH model geometry takes the form \cite{BH}:
\begin{equation}\label{metric function}
f(r)=1-\frac{2M}{r}+\frac{r^2}{L^2}+\alpha \bigg[\frac{\beta^2+3r^2+3\beta r}{3r(\beta+r)^3} \bigg].
\end{equation}
Here, $M$ illustrates the mass parameter of the BH. In this case, at large $r$, the asymptotic structure of this solution coincides with the standard AdS-Schwarzschild case, while the additional contribution proportional to $\alpha$ encodes corrections that affect the BH's physical properties. Also, this deformation of the AdS BH offers a useful framework for examining Newton-coupled extensions of GR in regimes of strong gravitational fields. In this context, the constants $\alpha$ and $\beta$ serve as controlling parameters that significantly modify several essential features of BH physics.  
The deformed Schwarzschild-AdS BH models described by the metric function $f(r)$ illustrate a generalized class of spherically symmetric solutions in which the standard Schwarzschild-AdS geometry is enriched by the presence of deformation parameters $\alpha$ and $\beta$, in addition to the conventional mass parameter $M$ and the AdS curvature radius $L$. Also, the first two terms, $1-2M/r$, encode the usual Schwarzschild contribution, where $M$ controls the gravitational mass and determines the location of the event horizon in the asymptotically flat case. Also, the term $r^2/L^2$ shows the influence of the cosmological constant $\Lambda=-3/L^2$, which introduces the AdS asymptotics, causing the spacetime to be confining at large $r$ in contrast to the flat Schwarzschild geometry. The last contribution, controlled by $\alpha$, is the deformation sector, which depends not only on $\alpha$ but also on the structural parameter $\beta$, and it modifies the geometry primarily at small and intermediate radial scales. When $\alpha=0$, the solution reduces smoothly to the Schwarzschild-AdS BH, ensuring consistency with known results. In the opposite regime, as $\beta\to\infty$, the deformation term vanishes as well, since the denominator $(\beta+r)^3$ suppresses the modification, thereby recovering again the Schwarzschild-AdS solution even for nonzero $\alpha$. In this case, for small $\beta$, the deformation becomes more pronounced near the horizon and in the near-singularity region, altering the effective gravitational potential and potentially shifting the horizon structure, the thermodynamic behavior, and the geodesic dynamics. Physically, the parameter $\alpha$ controls the “strength” of the deviation from the usual BH background, so that positive values enhance the effective potential barrier for scalar or particle propagation, while negative values would reduce it, leading to novel stability features. The interplay between $M$, $L$, $\alpha$, and $\beta$ therefore determines the geometry: in the large-$r$ limit, the $r^2/L^2$ term dominates, confirming that the spacetime remains asymptotically AdS regardless of deformation; in the NH regime, the competition between $-2M/r$ and the deformation term dictates horizon locations, which may shift compared to the undeformed case, thereby modifying the Hawking temperature and entropy. In this case, in the extreme small-$r$ limit, the deformation term behaves like $\sim \alpha/(3r\beta)$ for finite $\beta$, which effectively regularizes or modifies the singular behavior depending on the sign and magnitude of $\alpha$. Also, these deformed BH models provide a richer framework to probe how additional microscopic or effective-field-theory-motivated corrections, encoded in $\alpha$ and $\beta$, can influence both the gravitational background and the dynamical evolution of test fields such as the KG scalar, offering a versatile arena for studying modified gravity effects, AdS/CFT correspondence in non-standard geometries and the thermodynamic stability of BHs beyond the standard Schwarzschild-AdS models.

{The energy conditions provide restrictions on the source
supporting the deformed geometry. In units $G=c=1$, we write
the field equations as
\begin{equation}
G^\mu_\nu+\Lambda\delta^\mu_\nu
=8\pi T^\mu_\nu,
\qquad
\Lambda=-\frac{3}{L^2},
\end{equation}
where $T^\mu{}_\nu=\operatorname{diag}(-\rho,p_r,p_t,p_t)$
describes the deformation source. The cosmological constant
is retained on the geometric side of the field equations.
For the static, spherically symmetric metric considered here,
the independent components give
\begin{align}
8\pi\rho=\frac{1-f-rf'}{r^2}-\Lambda,
\quad
8\pi p_r=\frac{f-1+rf'}{r^2}+\Lambda,
\quad
8\pi p_t=\frac{f''}{2}+\frac{f'}{r}+\Lambda.
\end{align}
Substitution of the metric function yields \cite{BH}:
\begin{equation}
\rho=\frac{\alpha}{8\pi(r+\beta)^4},
\qquad
p_r=-\frac{\alpha}{8\pi(r+\beta)^4},
\qquad
p_t=\frac{\alpha(r-\beta)}{8\pi(r+\beta)^5}.
\label{eq:deformation_stress}
\end{equation}}

{We consider $\beta>0$, which avoids a pole at a positive
radius arising from $r+\beta=0$. The null energy condition
requires $\rho+p_r\geq0$ and $\rho+p_t\geq0$, whereas the
weak energy condition additionally requires $\rho\geq0$.
For the present source,
\begin{equation}
\rho+p_r=0,
\qquad
\rho+p_t=\frac{\alpha r}{4\pi(r+\beta)^5}.
\end{equation}
Consequently, both conditions hold for $\alpha\geq0$
and $\beta>0$. Negative values of $\alpha$ violate the
weak energy condition and the tangential null energy
condition on this branch. The dominant energy condition requires
$\rho\geq0$ and $\rho\geq|p_i|$. In addition to the
relations above, we obtain \cite{BH}:
\begin{equation}
\rho-p_r=\frac{\alpha}{4\pi(r+\beta)^4},
\qquad
\rho-p_t=\frac{\alpha\beta}{4\pi(r+\beta)^5}.
\end{equation}
Thus, the dominant energy condition is also satisfied
for $\alpha\geq0$ and $\beta>0$.}

{The strong energy condition requires
$\rho+p_i\geq0$ and $\rho+p_r+2p_t\geq0$. The latter
combination is
\begin{equation}
\rho+p_r+2p_t=\frac{\alpha(r-\beta)}{4\pi(r+\beta)^5}.
\end{equation}
For $\alpha>0$, this condition holds for $r\geq\beta$.
It is therefore satisfied throughout the exterior
$r\geq r_h$ if $r_h\geq\beta$. If $r_h<\beta$, it is
violated in the exterior interval $r_h<r<\beta$.
For $\alpha=0$, the deformation source vanishes and
all these conditions are saturated.
These statements refer to the deformation source rather
than to a stress tensor that includes the cosmological
constant. If the latter is treated as an effective fluid,
its energy density is $\rho_\Lambda=\Lambda/(8\pi)<0$.
The combined energy density then approaches
$\Lambda/(8\pi)$ at infinity, so the combined source
does not satisfy the weak energy condition throughout
the asymptotically AdS region. The energy conditions must be supplemented by the
requirement that the geometry possesses an event horizon ($r_h$).
The horizon equation gives
\begin{equation}
M=\frac{r_h}{2}\left(1+\frac{r_h^2}{L^2}\right)+\frac{\alpha(\beta^2+3\beta r_h+3r_h^2)}
{6(\beta+r_h)^3}.
\label{eq:ec_horizon_mass}
\end{equation}
Here, $r_h$ must be the largest positive root of $f(r)=0$,
with $f(r)>0$ outside it. Using this relation, the
Hawking temperature becomes
\begin{equation}
T_H=\frac{f'(r_h)}{4\pi}=\frac{1}{4\pi r_h}\left[1+\frac{3r_h^2}{L^2}
-\frac{\alpha r_h^2}{(\beta+r_h)^4}
\right].
\end{equation}
Hence, a nonextremal outer horizon requires
\begin{equation}
0\leq\alpha<
\frac{(\beta+r_h)^4}{r_h^2}
\left(1+\frac{3r_h^2}{L^2}\right),
\label{eq:ec_temperature_bound}
\end{equation}
together with Eq.~\eqref{eq:ec_horizon_mass}.
Equality corresponds to $f(r_h)=f'(r_h)=0$.
The energy conditions alone impose no upper bound on
$\alpha$; the additional restriction follows from
the horizon structure.}

To study the propagation of the massless scalar mode, we shall begin by deriving its EoM and then analyze it analytically and compute the \textsl{GBF}. Also, this prescribes the particles as having no extra interactions and only interacting weakly with gravity. In this context, based on such assumptions, the EoM can be written as
\begin{equation}\label{4}
\sqrt{-g}\nabla_{\beta}\nabla^{\beta}\Psi=\partial_{\beta}[\sqrt{-g}g^{\beta\gamma}\partial_{\gamma}\Psi]=0,
\end{equation}
here we can denote the $\Psi=\Psi(t,r,\theta,\phi)$ as the massless scalar field. Further, we can consider Eq. (\ref{metric function}) to obtain
\begin{eqnarray}\label{5}
\frac{-r^{2}\sin\theta}{f(r)}\partial_{tt}\Psi+\sin\theta
\partial_{r} \bigg(r^{2} f(r)\partial_{r}\Psi\bigg)+\partial_{\theta}
\bigg(\sin\theta\partial_{\theta}\Psi\bigg)+
\frac{1}{\sin\theta}\partial_{\phi\phi}\Psi =0.
\end{eqnarray}
By applying the technique of variable separation, it is possible to derive
\begin{equation}\label{6}
\Psi=\exp(-\iota w t)\exp(\iota m\phi)R_{w l m}(r)Y_l(\Omega),
\end{equation}
{We can specify the angular spheroidal function as $ Y_l(\Omega)$ denotes the 3-dimensional spherical harmonic function. In this case, using Eq. (\ref{5}), one can derive the radial part of the motion equations, as expressed in the following form \cite{GBF, GBF1}.
\begin{eqnarray}\label{7}
\frac{\partial }{\partial r}\bigg[r^{2}f(r)\frac{\partial R_{w l m}}{\partial
r}\bigg]+\bigg[\frac{r^{2}\omega^{2}}{f(r)}-
\lambda^{m}_{l}\bigg]R_{w l m}(r)=0.
\end{eqnarray}
The separation constant \(\lambda_{l}^{m}\) is related to the connection of the decoupled equations. It can be expressed as \cite{GBF, GBF1}.
\begin{equation}\label{9}
\lambda^{m}_{l}=(l^{2}+l).
\end{equation}
This constraint is essential in determining the appropriate radial solutions of the equation of motion (\ref{7}). In this case, such solutions provide access to the \textsl{GBF} associated with a massless scalar field. Furthermore, by employing the previously derived radial solution, it becomes possible to analyze in detail the structure of the effective potential near the \textsl{GBF}. To facilitate this investigation, a new form of radial transformation is introduced.
\begin{equation}\label{11}
R_{wlm}(r)=\frac{U_{wlm}(r)}{r}.
\end{equation}
By introducing the tortoise coordinate $x_{*}$, we obtain
\begin{equation}\label{12}
\frac{dx_{*}}{dr}=\frac{1}{f(r)},
\end{equation}
and
\begin{eqnarray}\label{13}
\frac{d}{dx_{*}}=f(r)\frac{d}{dr},
\end{eqnarray}
\begin{eqnarray}\label{14}
\frac{d^{2}}{dx_{*}^{2}}&=&f^{2}(r)\frac{d^{2}}{dr^2}+ff'\frac{d}{dr}.
\end{eqnarray}
As $r$ tends to infinity, the tortoise coordinate behaves as $x_{*}\to +\infty$, whereas when $r$ approaches the horizon radius $r_{h}$, it follows that $x_{*}\to -\infty$. Consequently, the coordinate $x_{\*}$ covers the entire interval from $-\infty$ to $+\infty$, in contrast to Eq.(\ref{7}), whose validity is limited to regions sufficiently far from the BH horizon. Also, in this formulation, Eq.(\ref{7}) is interpreted as a Schrödinger-type wave equation.
\begin{equation}\label{15}
\left(\frac{d^2}{dx_{*}^2}+\omega^2-V_{eff}\right)U_{wlm}(r)=0.
\end{equation}
The effective potential's corresponding form results in
\begin{eqnarray}\nonumber
V_{eff}=\frac{ff'}{r}+\frac{\lambda_{l}^{m}f(r)}{r^2}.
\end{eqnarray}

For different values of the deformation parameter, the effective potential as a function of $r_h$ is visually investigated. Figs. (\ref{v1})–(\ref{v3}) show the method by which the BH's mass, length parameter and deformation parameters affect the effective potential. In this case, as seen in Fig. (\ref{v1}), a greater peak in the potential function results from increasing the deformation parameter $\alpha$. It is found that it decreases by increasing the parameter $\beta$. In contrast, as seen in Fig. (\ref{v2}), the cosmological length parameter also affects the behavior of the effective potential. It is found that the potential function decreases by increasing $L$ as shown in Fig. (\ref{v2}). Further, we found a similar behavior for the mass of BH. By enhancing the mass of BH, the effective potential decreases as shown in  Fig. (\ref{v3}).
\begin{figure}\center
\epsfig{file=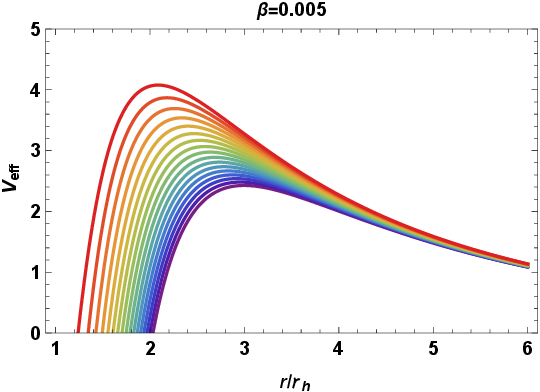,width=.4\linewidth,height=2.2in}\quad\epsfig{file=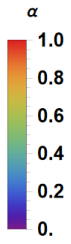,width=.06\linewidth,height=2.1in}\quad\epsfig{file=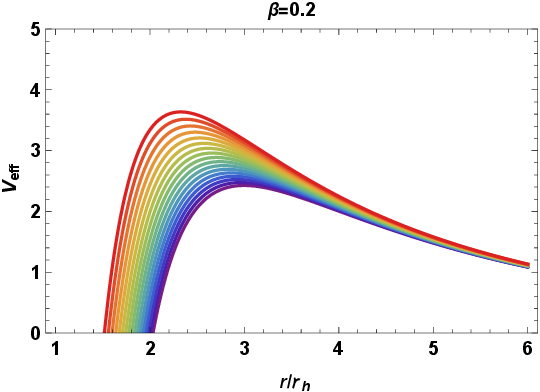,width=.4\linewidth,height=2.2in}\quad\epsfig{file=al.eps,width=.06\linewidth,height=2.1in}
\caption{\label{v1}{$V_{eff}$ versus $r_h$ by varying $\alpha$ and $\beta$ with $l=8,M=1,L=10$.}}
\epsfig{file=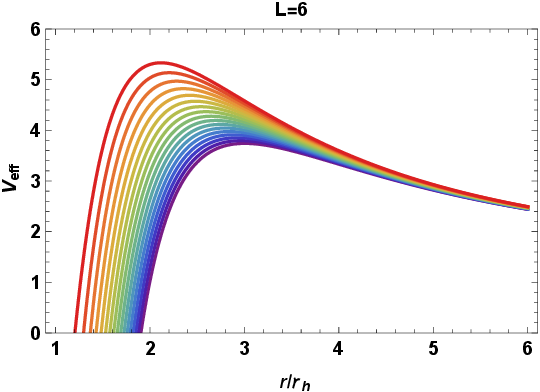,width=.4\linewidth,height=2.2in}\quad\epsfig{file=al.eps,width=.06\linewidth,height=2.1in}\quad\epsfig{file=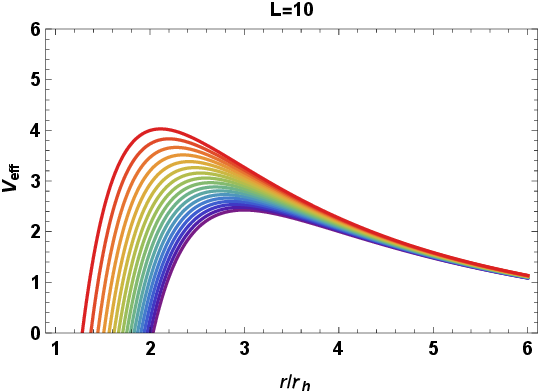,width=.4\linewidth,height=2.2in}\quad\epsfig{file=al.eps,width=.06\linewidth,height=2.1in}
\caption{\label{v2}{$V_{eff}$ versus $r_h$ by varying $\alpha$ and $L$ with $l=10,M=1, \beta=0.08$.}}
\epsfig{file=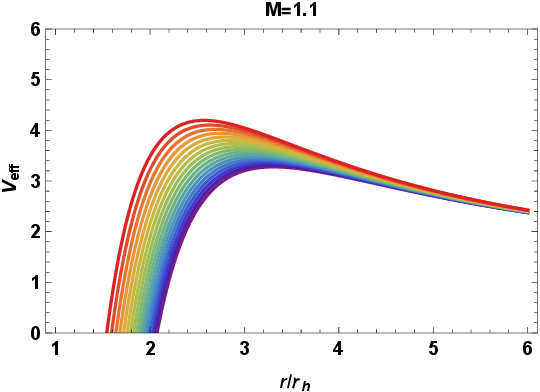,width=.4\linewidth,height=2.2in}\quad\epsfig{file=al.eps,width=.06\linewidth,height=2.1in}\quad\epsfig{file=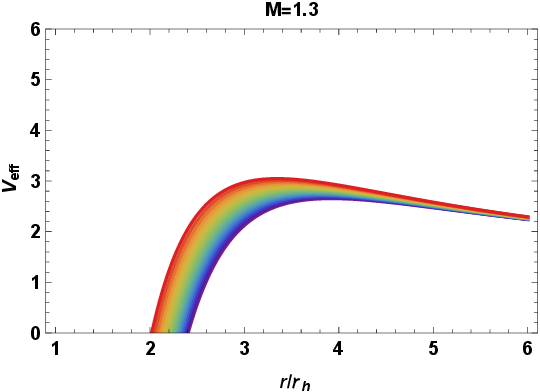,width=.4\linewidth,height=2.2in}\quad\epsfig{file=al.eps,width=.06\linewidth,height=2.1in}
\caption{\label{v3}{$V_{eff}$ versus $r_h$ by varying $\alpha$ and $M$ with $L=20, \beta=0.08, l=3$.}}
\end{figure}

\section{Greybody Factor}\label{S3}

The radial Schrödinger EoM (\ref{7}) has analytic solutions to the \textsl{GBF} and these are further ahead in this section. Our two asymptotic solutions correspond to the different regimes, or rather to the ones around the BH horizon and far away. The solutions are perfectly combined in a middle domain to develop a solution that is valid in the whole domain. After using such a transformation as $r_h$, we get the NH analytical expression
\begin{eqnarray}\label{16}
r\rightarrow
S=\frac{1-\frac{2M}{r}+\frac{r^2}{L^2}+\alpha \bigg[\frac{\beta^2+3r^2+3\beta r}{3r(\beta+r)^3} \bigg]}{1+\frac{r^2}{L^2}+\alpha \bigg[\frac{\beta^2+3r^2+3\beta r}{3r(\beta+r)^3} \bigg]},
\end{eqnarray}
which yields
\begin{eqnarray}\nonumber
\frac{d S}{d r}=\frac{(1-S)U(r_{h})}{r_{h}}.
\end{eqnarray}
Hence, we have
\begin{eqnarray}\nonumber
U(r_{h})= \frac{L^2(1+O+r O')+3r^2}{L^2+r^2+O L^2}, \quad
O'(r_{h})= -\frac{\alpha \bigg[6r^2(\beta+r)+\beta^2 (4r+\beta)\bigg]}{3r^2 (\beta+r)^4}
\end{eqnarray}
Based on these findings, the radial equation is represented in the form (\ref{7}).
\begin{equation}\label{17}
S(1-S)\frac{d^{2}R_{w l m}}{d S^{2}}+(A-BS)\frac{d R_{w l m}}{d S}+
\frac{1}{(1-S)U^{2}}\bigg[\frac{\chi_{h}}{S}-\lambda_{h}\bigg]R_{w l
m}=0,
\end{equation}
where
\begin{eqnarray}\nonumber
A=\frac{r_h}{(1-S)U(r_{h})}\frac{d}{dr}(S-S^{2}),\quad
B=-\frac{S}{U^{2}(r_{h})\bigg(1+\frac{r^2}{L^2}+\alpha \bigg[\frac{\beta^2+3r^2+3\beta r}{3r(\beta+r)^3} \bigg]\bigg)}\frac{d}{dr}\bigg[r_{h}U(r_{h}) \bigg(1+\frac{r^2}{L^2}+\alpha \bigg[\frac{\beta^2+3r^2+3\beta r}{3r(\beta+r)^3} \bigg]\bigg)\bigg],\\\nonumber
\end{eqnarray}

\begin{eqnarray}\nonumber
\chi_{h}&=&\frac{w^{2}r^{2}_{h}}{\bigg(1+\frac{r^2}{L^2}+\alpha \bigg[\frac{\beta^2+3r^2+3\beta r}{3r(\beta+r)^3} \bigg]\bigg)^2},\quad \lambda_{h}= \frac{\lambda^{m}_{l}}{1+\frac{r^2}{L^2}+\alpha \bigg[\frac{\beta^2+3r^2+3\beta r}{3r(\beta+r)^3} \bigg]}.
\end{eqnarray}

The function presented in Eq. (\ref{17}) can be reformulated as
\begin{equation}\label{18}
R_{w l m}(S)=S^{\epsilon_{1}}(1-S)^{\eta_{1}}\hat{F}(S).
\end{equation}
Also, Eq.(\ref{17}) is defined as follows:
\begin{eqnarray}\nonumber
S(1-S)\frac{d^{2}\hat{F}(S)}{d S^{2}} +
\bigg[2\epsilon_{1}+A-(2\epsilon_{1}+2\eta_{1}+B)S\bigg]\frac{d
\hat{F}(S)}{d S}+ \bigg[\bigg(\epsilon^{2}_{1}-\epsilon_{1}\\+
A\epsilon_{1}+\frac{\chi_{h}}{U^{2}} \bigg)\frac{1}{S}+\label{18a}
\bigg(\eta^{2}_{1}-\eta_{1}-\eta_{1}A+\eta_{1}B+
\frac{\chi_{h}}{U^{2}}-\frac{\lambda_{h}}{U^{2}}\bigg)\frac{1}
{1-S}\bigg]\hat{F}(S)=0.
\end{eqnarray}
One possible approach to determine the power coefficients $\epsilon_{1}$ and $\eta_{1}$ is by
\begin{eqnarray}\nonumber
\epsilon^{2}_{1}-\epsilon_{1}+A\epsilon_{1}+\frac{\chi_{h}}
{U^{2}}=0,\\\nonumber
\eta^{2}_{1}-\eta_{1}-\eta_{1}A+\eta_{1}B+\frac{\chi_{h}}
{U^{2}}-\frac{\lambda_{h}}{U^{2}}=0.
\end{eqnarray}
In this case, by solving Eq.(\ref{18a}), one arrives at the hypergeometric (HG) differential equation expressed as
\begin{equation}\label{19}
S(1-S)\frac{d^{2} \hat{F}(S)}{d
S^{2}}+\bigg[\bar{c}_{1}-(\bar{a}_{1}+\bar{b}_{1}+1)S\bigg]\frac{d
\hat{F}(S)}{d S}-\bar{a}_{1}\bar{b}_{1}\hat{F}(S)=0,
\end{equation}
with
\begin{eqnarray}\nonumber
\bar{a}_{1}=\eta_{1}+\epsilon_{1}+B-1,\quad\bar{b}_{1}=\eta_{1}+
\epsilon_{1},\quad\bar{c}_{1}= 2\epsilon_{1}+A.
\end{eqnarray}
For the NH, its general solution can be displayed as
\begin{eqnarray}\label{20}
(R_{w l m})_{NH}(S)=
\hat{A_{1}}S^{\epsilon_{1}}(1-S)^{\eta_{1}}\hat{F}(\bar{a}_{1},\bar{b}_{1},\bar{c}_{1}
; S )+\hat{A_{2}}S^{-\epsilon_{1}}(1-S)^{\eta_{1}}
\hat{F}(1-\bar{c}_{1}+\bar{a}_{1},
1+\bar{b}_{1}-\bar{c}_{1},-\bar{c}_{1}+2;S),
\end{eqnarray}
with $\hat{A_{1}}$ and $\hat{A_{2}}$ treated as fixed constants
\begin{eqnarray}\nonumber
\epsilon^{\pm}_{1}&=& \frac{1}{2}\bigg[(1-A)\pm\sqrt{(1-A)^{2}-4
\frac{\chi_{h}}{U^{2}}}\bigg],\\\nonumber
\eta^{\pm}_{1}&=&\frac{1}{2}\bigg[(1+A-B)\pm\sqrt{(1+A-B)^{2}+4\bigg(
\frac{\lambda_{h}}{U^{2}}-\frac{\chi_{h}}{U^{2}}\bigg)}\bigg].
\end{eqnarray}
Under consideration of boundary conditions, it is seen that there are no outgoing modes in the dependence on the BH horizon. So, we are at liberty to make the further decision that $\hat{A}_{1} = 0$; or, if $\hat{A}_{1} \neq 0$, in such case we find that $\hat{A}_{2} = 0$ when we make the decision that $\epsilon_{1} = 0$. Note that $\hat{A}_{1}$ and $\hat{A}_{2}$ are independent of $\epsilon_{1}$ because it has two values, so on setting $\hat{A}_{2} = 0$, we find that $\epsilon_{1}^{+} = \epsilon_{1}^{-}$. The convergence condition of the HG function is the other choice to calculate the parameters of $\eta_{1}$, to be applied in the case of $\eta_{1}^{+} = \eta_{1}^{-}$. In turn, the analytic version of the NH solution is thus expressed as
\begin{equation}\label{21}
(R_{w l m})_{N H}(S)=\hat{A_{1}}S^{\epsilon_{1}}(1-S)^{\eta_{1}}
\hat{F}(\bar{a}_{1}, \bar{b}_{1},\bar{c}_{1};S).
\end{equation}

Substituting $f(r)$ with $T(r)$ allows us to determine the form of the radial equation associated with the far-region BH horizon, following the same method employed for the NH case 
\begin{equation}\label{22}
T(r)=\frac{1+\frac{r^2}{L^2}+\alpha \bigg[\frac{\beta^2+3r^2+3\beta r}{3r(\beta+r)^3} \bigg]}{r^2}.
\end{equation}
As a result, the modification appears in the form of the radial equation.
\begin{eqnarray}\nonumber
&&T(1-T)\frac{d^{2} R_{ w l m}}{dT^{2}}+(C-D^{\ast}T)\frac{d  R_{ w
l m}}{d T}+\frac{1}{D^2(1-T)}\bigg[\frac{\chi_{f}}{T}-\lambda_{f}
\bigg] R_{ w l m}=0,
\end{eqnarray}
where
\begin{eqnarray}\nonumber
\chi_{f}&=&\left(\frac{r w T(1-T)}{f(r)}\right)^2,\quad
\lambda_{f}=\frac{\lambda_{l}^{m}T(1-T)^2}{f(r)},\\\nonumber \quad
C&=&\frac{1+\frac{r^2}{L^2}+\alpha \bigg[\frac{\beta^2+3r^2+3\beta r}{3r(\beta+r)^3} \bigg]}{r D^2 f(r)} \frac{d}{dr}\bigg[D f(r)\bigg],\quad
D^{\ast}=\frac{1+\frac{r^2}{L^2}+\alpha \bigg[\frac{\beta^2+3r^2+3\beta r}{3r(\beta+r)^3} \bigg]}{r^2 D},
\\\nonumber \quad D&=& \frac{1}{r}\frac{d}{dr}\bigg(1+\frac{r^2}{L^2}+\alpha \bigg[\frac{\beta^2+3r^2+3\beta r}{3r(\beta+r)^3} \bigg]\bigg)-\frac{2}{r^2}\bigg(1+\frac{r^2}{L^2}+\alpha \bigg[\frac{\beta^2+3r^2+3\beta r}{3r(\beta+r)^3} \bigg]\bigg)
\end{eqnarray}
A redefinition of the field within the previously stated differential equation is required as
\begin{eqnarray}\label{23}
R_{wlm}(T)=T^{\epsilon_{2}}(1-T)^{\eta_{2}}\hat{F(T)},
\end{eqnarray}
we get
\begin{eqnarray}\nonumber
&&T(1-T)\frac{d^{2}\hat{F(T)}}{d T^{2}}+\bigg[ 2\epsilon_{2} + C -
(2\epsilon_{2}+2\eta_{2} + D^{\ast})T\bigg]\frac{d \hat{F(T)}}{d T}
+\bigg[\bigg(\epsilon^{2}_{2}-\epsilon_{2}\\&&+\epsilon_{2}C+
\frac{\chi_{f}}{D^2}\bigg)\bigg(
\eta_{2}^{2}-\eta_{2}-\eta_{2}C+\eta_{2} D^{\ast} +
\frac{\chi_{f}}{D^2}-\frac{\lambda_{f}}{D^2}
\bigg)\frac{1}{1-T}\bigg]\hat{F(T)}=0.\nonumber
\end{eqnarray}
The values of the power coefficients $\epsilon_{2}$ and $\eta_{2}$ are determined as
\begin{eqnarray}\nonumber
\epsilon^{2}_{2}-(1-C)\epsilon_{2}+\frac{\chi_{f}}{D^2}=0,\\\nonumber
\eta^{2}_{2}-(1-D^{\ast}+C)\eta_{2}+\frac{\chi_{f}}
{D^2}-\frac{\lambda_{f}}{D^2}=0.
\end{eqnarray}
Hence, the HG-based equation is computed as
\begin{eqnarray}\nonumber
\bar{a}_{2}=\epsilon_{2}+\eta_{2}+D^{\ast}-1,\quad\bar{b}_{2}=
\epsilon_{2}+\eta_{2},\quad\bar{c}_{2}= 2\epsilon_{2}+C,
\end{eqnarray}
which becomes
\begin{equation}\nonumber
T(1-T)\frac{d^{2}\hat{F(T)}}{d T^{2}}+
\bigg[\bar{c}_{2}-(1+\bar{a}_{2}+\bar{b}_{2})\bigg]\frac{d
\hat{F(T)}}{d T}-\bar{a}_{2}\bar{b}_{2} \hat{F(T)}=0.
\end{equation}
Its general solution is
\begin{eqnarray}\label{24}
(R_{wlm})_{f}(T)=
\hat{B_{1}}T^{\epsilon_{2}}(1-T)^{\eta_{2}}\hat{F}(\bar{a}_{2},
\bar{b}_{2},\bar{c}_{2};T)+\hat{B_{2}}T^{-\epsilon_{2}}
(-T+1)^{\eta_{2}}
\hat{F}(1-\bar{c}_{2}+\bar{a}_{2},1-\bar{c}_{2}+\bar{b}_{2},
-\bar{c}_{2}+2;T),
\end{eqnarray}
with
\begin{eqnarray}\nonumber
\epsilon_{2}^{\pm}&=& \frac{1}{2}\bigg[(1-C)\pm \sqrt{(1-C)^{2}-
4\frac{\chi_{f}}{D^2}}\bigg],\\\nonumber \eta_{2}^{\pm}&=&
\frac{1}{2}\bigg[(1-D^{\ast}+C)\pm \sqrt{(1-D^{\ast}+C)^{2} -
4\bigg(\frac{\chi_{f}}{D^2}
-\frac{\lambda_{f}}{D^2}\bigg)}\bigg],
\end{eqnarray}
where the constants $\hat{B_{1}}$ and $\hat{B_{2}}$ are randomly chosen. Identical to NH, we compute the solutions through
the HG convergence constraint, in which
$\epsilon_{2}^{-}=\epsilon_{2}^{+}$ and $\eta_{2}^{+}=\eta_{2}^{-}$
are similar.

\section{Matching with Intermediate Zone}\label{S4}

In this section, we aim to smoothly connect the “NH” solution with the “faraway” solution within the intermediate region for all values of $r$. To extend the scope of the analysis, we substitute the HG function parameter $S$ in Eq. (\ref{21}) with $1-S$. As a result, we obtain
\begin{eqnarray}\nonumber
(R_{w l m})_{N H}(S)&=& (-S+1)^{\eta_{1}}
\bigg[\frac{\Gamma(\bar{c}_{1})\Gamma(\bar{c}_{1}-\bar{a}_{1}-\bar{b}_{1})
}{\Gamma(-\bar{a}_{1}+\bar{c}_{1})
\Gamma(\bar{c}_{1}-\bar{b}_{1})} \hat{F}(\bar{a}_{1},\bar{b}_{1}
,\bar{c}_{1};1-S)+(1-S)^{\bar{c}_{1}-\bar{a}_{1}-\bar{b}_{1}}
\frac{\Gamma(\bar{c}_{1})
\Gamma(-\bar{c}_{1}+\bar{b}_{1}+\bar{a}_{1})}{\Gamma(\bar{b}_{1})
\Gamma(\bar{a}_{1})}
\\\label{25}&\times& \hat{F}(+\bar{c}_{1}-\bar{a}_{1},
+\bar{c}_{1}-\bar{b}_{1},1-\bar{a}_{1}-\bar{b}_{1}+\bar{c}_{1} ; 1-S
)\bigg]\hat{A_{1}}S^{\epsilon_{1}}.
\end{eqnarray}
Through Eq. (\ref{16}), we get
\begin{eqnarray}\nonumber
1-S = \frac{2ML^2}{r(L^2+r^2+O L^2)}.
\end{eqnarray}
For the case where $f(r)=0$, considering the asymptotic limit $r\gg r_{h}$ and taking $S\rightarrow1$ with $2M$ as the corresponding value, the location of the horizon can be determined accordingly. Also, under these conditions, the stretched NH geometry takes the form 
\begin{eqnarray}\nonumber
(1-S)^{\eta_{1}}\simeq
\bigg[\frac{r_{h}}{r}\frac{1+L^{-2}_{\ast}}{1+ r^2 L^{-2}_{\ast}}\bigg]^{\eta_{1}}\\\nonumber
&\simeq & \bigg[\frac{r_{h}}{r}\frac{1+L^{-2}_{\ast}}{1+ r^2 L^{-2}_{\ast}}\bigg]^{-l},
\end{eqnarray}
and
\begin{eqnarray}\nonumber
(1-S)^{\eta_{1}-\bar{a}_{1}-\bar{b}_{1}+\bar{c}_{1}}&\simeq &
\bigg[\frac{r_{h}}{r}\frac{1+L^{-2}_{\ast}}{1+ r^2 L^{-2}_{\ast}}\bigg]^{\eta_{1}}\\\nonumber
&\simeq & \bigg[\frac{r_{h}}{r}\frac{1+L^{-2}_{\ast}}{1+ r^2 L^{-2}_{\ast}}\bigg]^{-\eta_{1}+A-B+1} \\\nonumber
&\simeq & \bigg[\frac{r_{h}}{r}\frac{1+L^{-2}_{\ast}}{1+ r^2 L^{-2}_{\ast}}\bigg]^{1+l},
\end{eqnarray}
with $L_{\ast}=\tfrac{L}{r_{h}}$ accordingly. Also, during the intermediate stage, the NH configuration takes the form of
\begin{equation}\label{26}
(R_{w l m})_{NH}(S)=\tilde{A_{1}}\bigg(\frac{r}{r_{h}}
\bigg)^{l}+\tilde{A_{2}}\bigg(\frac{r}{r_{h}}\bigg)^{-(1+l)},
\end{equation}
with
\begin{eqnarray}\nonumber
\tilde{A_{1}}&=&\hat{A_{1}}\bigg[\frac{1+L^{-2}_{\ast}}{1+ r^2 L^{-2}_{\ast}}\bigg]^{-l}
\frac{\Gamma(\bar{c}_{1})\Gamma(-\bar{a}_{1}-\bar{b}_{1}+\bar{c}_{1})
}{\Gamma(\bar{c}_{1}-\bar{b}_{1})\Gamma(\bar{c}_{1}-
\bar{a}_{1})},\\\nonumber \tilde{A_{2}}&=&\hat{A_{1}}
\bigg[\frac{1+L^{-2}_{\ast}}{1+ r^2 L^{-2}_{\ast}}\bigg]^{1+l}
\frac{\Gamma(\bar{a}_{1}+\bar{b}_{1}-\bar{c}_{1})
\Gamma(\bar{c}_{1})}{\Gamma(\bar{a}_{1})\Gamma(\bar{b}_{1})}.
\end{eqnarray}

By substituting $T$ with $1-T$, the HG function parameters are modified. For lower values of $\alpha$ and $\beta$, they are limited as $T(r_f) \to 0$. This change offers a solution that extends much beyond the event horizon of the BH. Therefore, Eq. (\ref{22}) reduces to:
\begin{eqnarray}\nonumber
(1-T)^{\eta_{2}}\simeq \bigg[\frac{r^2-r^{2}_{f}}{r r^{3}_{f}}\bigg]^{-l}
\bigg(\frac{r}{r_{f}} \bigg)^{l},
\end{eqnarray}
and
\begin{eqnarray}\nonumber
(1-T)^{\eta_{2}+\bar{c}_{2}-\bar{a}_{2}-\bar{b}_{2}}\simeq\bigg[\frac{r^2-r^{2}_{f}}{r r^{3}_{f}}\bigg]^{1+l}
\bigg(\frac{r}{r_{f}}\bigg)^{-(1+l)}.
\end{eqnarray}
The outcome obtained from Eq.(\ref{24}) in the regime associated with the distant horizon results in
\begin{equation}\label{27}
(R_{w l m})_{f}(T)=\bigg( \tilde{H_{1}}\tilde{B_{1}}+ \tilde{H_{3}}
\tilde{B_{2}}\bigg)\bigg(\frac{r}{r_{f}}\bigg)^{l}+\bigg(
\tilde{H_{2}}\tilde{B_{1}}+
\tilde{B_{2}}\tilde{H_{4}}\bigg)\bigg(\frac{r}{r_{f}}\bigg)^{-(1+l)},
\end{equation}
where
\begin{eqnarray}\nonumber
\tilde{H_{1}}&=&\frac{\Gamma(\bar{c}_{2})
\Gamma(\bar{c}_{2}-\bar{a}_{2}-\bar{b}_{2})}
{\Gamma(\bar{c}_{2}-\bar{a}_{2})\Gamma(\bar{c}_{2}-\bar{b}_{2})}
\bigg[\frac{r^2-r^{2}_{f}}{r r^{3}_{f}}\bigg]^{-l},\\\nonumber\tilde{H_{2}}&=&
\frac{\Gamma(\bar{c}_{2})\Gamma(-\bar{c}_{2}+\bar{b}_{2}+\bar{a}_{2})
} { \Gamma(\bar{b}_{2})\Gamma(\bar{a}_{2})}\bigg[\frac{r^2-r^{2}_{f}}{r r^{3}_{f}}\bigg]^{1+l},\\\nonumber\tilde{H_{3}}
&=& \frac{\Gamma(2-\bar{c}_{2})
\Gamma(\bar{c}_{2}-\bar{a}_{2}-\bar{b}_{2})}{\Gamma(1-\bar{a}_{2})
\Gamma(1-\bar{b}_{2})} \bigg[\frac{r^2-r^{2}_{f}}{r r^{3}_{f}}\bigg]^{-l},\\\nonumber\tilde{H_{4}}
&=& \frac{\Gamma(2-\bar{c}_{2})
\Gamma(\bar{a}_{2}+\bar{b}_{2}-\bar{c}_{2})}
{\Gamma(1+\bar{a}_{2}-\bar{c}_{2})
\Gamma(1+\bar{b}_{2}-\bar{c}_{2})}\bigg[\frac{r^2-r^{2}_{f}}{r r^{3}_{f}}\bigg]^{l+1}.
\end{eqnarray}

The asymptotic behaviors are analyzed by employing equivalent powers of $1+l$ and $l$ for a direct comparison between the two solutions.
\begin{eqnarray}\nonumber
\tilde{A_{1}}=\tilde{H_{3}}
\tilde{B_{2}}+\tilde{H_{1}}\tilde{B_{1}},\quad\tilde{A_{2}}=\tilde{B_{1}}\tilde{H_{2}} +
\tilde{H_{4}}\tilde{B_{2}} .
\end{eqnarray}
Additionally, $\tilde{B_{1}}$ and $\tilde{B_{2}}$ are the integrated constants. It can be expressed as
\begin{eqnarray}\nonumber
\tilde{B_{1}}=\frac{\tilde{A_{1}}
\tilde{H_{4}}-\tilde{A_{2}}\tilde{H_{3}}}{\tilde{H_{1}}
\tilde{H_{4}}-\tilde{H_{2}}\tilde{H_{3}}}\quad \tilde{B_{2}}=\frac{\tilde{A_{1}}\tilde{H_{2}}-\tilde{A_{2}}
\tilde{H_{1}}}{\tilde{H_{2}}\tilde{H_{3}}-\tilde{H_{1}}
\tilde{H_{4}}}.
\end{eqnarray}
The emission rate associated with a massless scalar field can be obtained by expressing the \textsl{GBF} in the following manner \cite{GBF, GBF1}.
\begin{eqnarray}\nonumber
|A_{l,m}|^{2} =
1-\bigg|\frac{\tilde{B_{2}}}{\tilde{B_{1}}}\bigg|^{2}=
1-\bigg|\frac{\tilde{A_{1}}\tilde{H_{2}}-\tilde{A_{2}}
\tilde{H_{1}}}{\tilde{A_{2}}\tilde{H_{3}}-\tilde{A_{1}}
\tilde{H_{4}}}\bigg|^{2}.
\end{eqnarray}

\begin{figure}\center
\epsfig{file=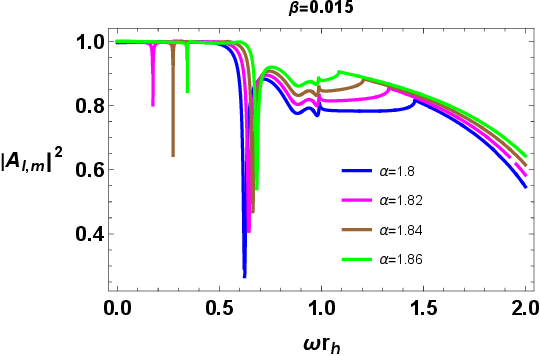,width=.45\linewidth,height=2.5in}\quad\quad\epsfig{file=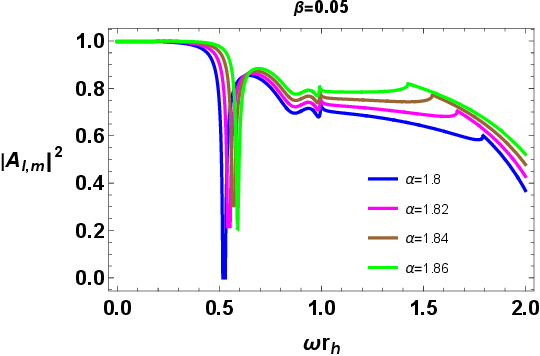,width=.45\linewidth,height=2.5in}
\caption{\label{gb1}{\textsl{GBF} versus $wr_{h}$ with different values of $\alpha$ and $\beta$ with $M=1,L=2,l=2$.}}
\epsfig{file=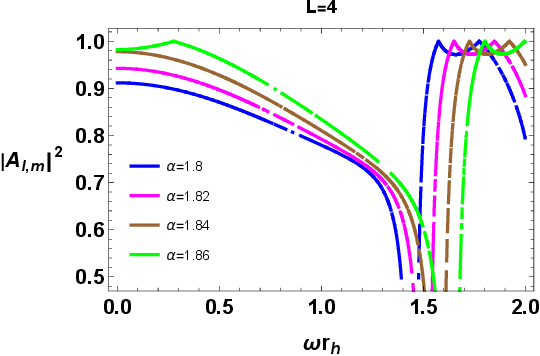,width=.45\linewidth,height=2.5in}\quad\quad\epsfig{file=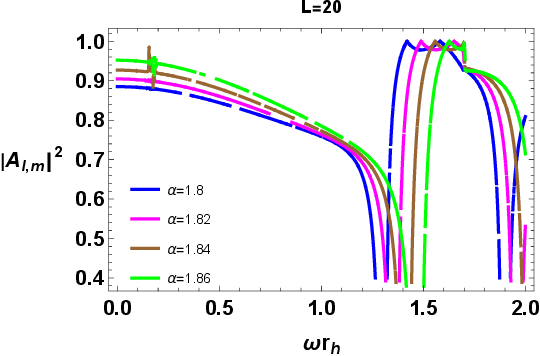,width=.45\linewidth,height=2.5in}
\caption{\label{gb2}{\textsl{\textsl{GBF}} versus $wr_{h}$ with different values of of $\alpha$ and $L$ with $M=1,\beta=0.05,l=2$.}}
\epsfig{file=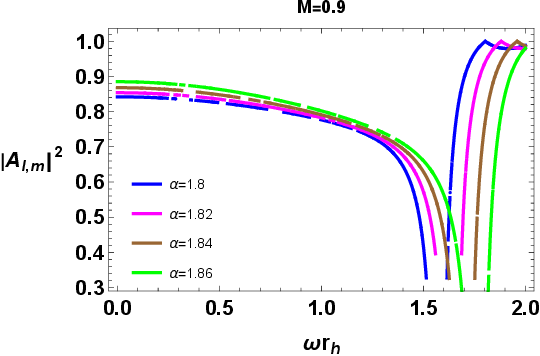,width=.45\linewidth,height=2.5in}\quad\quad\epsfig{file=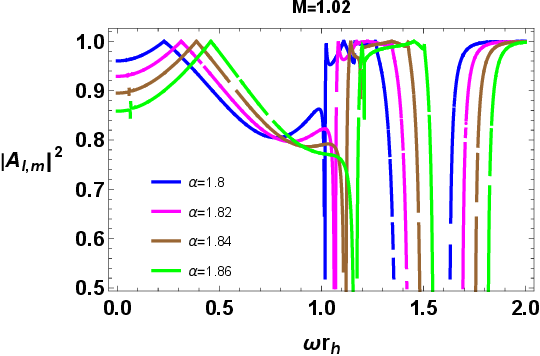,width=.45\linewidth,height=2.5in}
\caption{\label{gb3}{\textsl{GBF} versus $wr_{h}$ with different values of $\alpha$ and $M$ with $L=10,\beta=0.05,l=2$.}}
\end{figure}
As a visual inspection shows, wave modes travel across the event horizon well beyond the remote cosmic horizon. There is a particular relationship between the wave frequency and emission rate: the waves with low frequencies are usually reflected and the waves with higher frequencies are mostly passed. In general, the high-frequency waves are more efficiently able to penetrate potential obstacles. The influence of different physical parameters on the \textsl{GBF} in the range of absorption of massless scalar fields as well as their rate of emission is investigated using graphical techniques. Far in space, waves meet a potential object in their way that will either redirect them or pass them through. It is important that the frequency of waves is related to the effective potential. When the frequency of the wave has a higher energy than the effective potential, it passes through the barrier. If the potential is higher than the frequency, partial transmission as well as reflection lead to reduced \textsl{GBF}. It is observed that the absorption rate depends on the deformation parameters, BH mass, and $L$. There are different settings in which this scenario can be analyzed that could help to consider the numerical behavior of the \textsl{GBF}. The scattering and absorption of the \textsl{GBF} are used to measure particles that get close to a BH. It illustrates that a particle can enter the BH by absorption rather than by reflection or transmission. This is a fundamental concept to understanding BH thermodynamics, as well as the subsequent investigation of QFT in curved spacetime. The \textsl{GBF} shows that any potential barriers can cause changes in outgoing radiation; thus, it is an important tool in studying scalar field BH phenomena. This can improve our understanding of the rates of energy emission and the knowledge loss question. Radial oscillation in the rate of absorption of the scalar field can be seen in Figs. (\ref{gb1})-(\ref{gb3}) and as one can see, generally the rate of absorption decreases with the increase in frequency (see Fig. \ref{gb1}). When the deformation parameter $\alpha$ increases, the representation shows a slower absorption rate and as the frequency becomes high with increased values of $L$, the rate increases. An increasing value of M will show changes in the absorption rate at higher frequencies. The deformation parameter, the mass of the BH and $L$ are very sensitive to the rate of absorption.
\section{Joule-Thomson expansion: a thorough investigation}\label{S5}

In this part, setting the metric potential to vanish, the condition $f(r_{h})=0$ provides the value of $r_{h}$ which defines the location of the BH event horizon. 
We now use $P=-\Lambda/8\pi$ to obtain the cosmological constant, which represents the system's pressure. The respective expression of pressure becomes 
\begin{equation}\label{2-7}
P=\frac{3}{8\pi L^{2}}.
\end{equation}
Now, we can determine the entropy in terms of the BH horizon radius as
\begin{equation}\label{2-8}
S=\pi r^2_h.
\end{equation}
We analyze the BH's mass $M$, or $M=M(\alpha,\beta,P,S)$, based on entropy, deformation parameters and pressure   (see Equations (\ref{2-7}), and (\ref{2-8})). The mass $M$ of the BH therefore yields
\begin{equation}\label{2-9}
M=\frac{1}{6} \left(\frac{8 P S^{3/2}}{\sqrt{\pi }}+\frac{\alpha
\big(\beta ^2+\frac{3 \beta  \sqrt{S}}{\sqrt{\pi }}+\frac{3 S}{\pi
}\big)}{\big(\beta +\frac{\sqrt{S}}{\sqrt{\pi }}\big)^3}+\frac{3
\sqrt{S}}{\sqrt{\pi }}\right).
\end{equation}

{Considering the first law of BH thermodynamics given by
\begin{equation}\label{2-10}
dM=TdS+VdP,
\end{equation}
the respective thermodynamic parameters are expressed as
\begin{equation}\label{2-11}
T=\Big(\frac{\partial M}{\partial
S}\Big)_{P},~~V=\Big(\frac{\partial M}{\partial
P}\Big)_{S}.
\end{equation}
Therefore, using Eqs. (\ref{2-9}) and (\ref{2-11}), we obtain the BH temperature as entropy and pressure, which are provided by
\begin{eqnarray}\nonumber
T&=& \frac{1}{4 \sqrt{\pi } \sqrt{S} \big(\sqrt{\pi } \beta
+\sqrt{S}\big)^4}\bigg[\pi ^2 \beta ^4+4 \sqrt{\pi } \beta  S^{3/2}
\big(8 \pi \beta ^2 P+1\big)+32 \sqrt{\pi } \beta  P S^{5/2}+8 P
S^3+S^2 \big(48 \pi \beta ^2 P+1\big)\\\label{2-12}&+&\pi  S
\big(-\alpha+6 \beta ^2+8 \pi \beta ^4 P\big)+4 \pi ^{3/2} \beta ^3
\sqrt{S}\bigg].
\end{eqnarray}
Also, the BH volume $V$ can be expressed as
\begin{equation}\label{2-13}
V=\frac{4 S^{3/2}}{3 \sqrt{\pi }}.
\end{equation}}

Additionally, using the thermodynamic pressure (\ref{2-7}), entropy (\ref{2-8}) and temperature (\ref{2-12}), we can determine the BH's equation of state (EoS) as
\begin{equation}\label{2-16}
P=\frac{1}{8} \left(\frac{\frac{\alpha }{(\beta
+r_{h})^4}-\frac{1}{r_{h}^2}}{\pi }+\frac{4 T}{r_{h}}\right).
\end{equation}
To identify the critical points, one needs to determine the solutions of the following system.
\begin{equation}\label{2-17}
\frac{\partial P}{\partial r_{h}}=\frac{{\partial}^{2}
P}{{\partial}{r^2_{h}}}=0.
\end{equation}

\begin{figure}\center
\epsfig{file=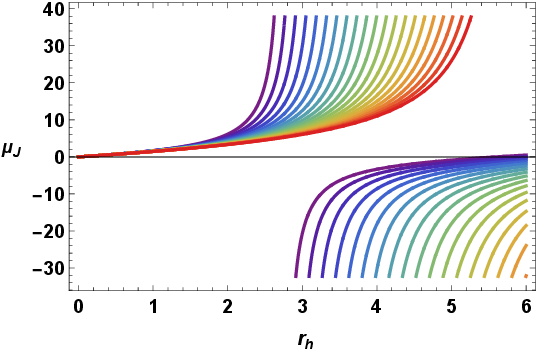,width=.4\linewidth,height=2.2in}
\quad\epsfig{file=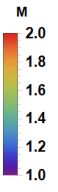,width=.06\linewidth,height=2.1in}
\quad\epsfig{file=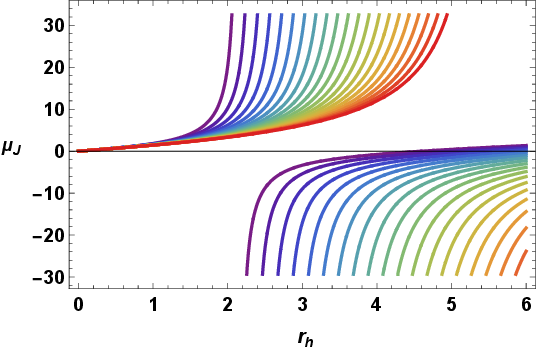,width=.4\linewidth,height=2.2in}
\quad\epsfig{file=m.eps,width=.06\linewidth,height=2.1in}
\caption{\label{jte1} Behavior of JT coefficient by varying $M$ with $\alpha=0.5$ (left plot) and $\alpha=1.5$ (right plot) with $\beta=0.05$}
\epsfig{file=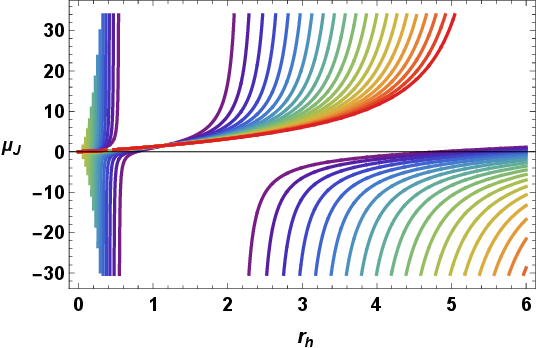,width=.4\linewidth,height=2.2in}
\quad\epsfig{file=m.eps,width=.06\linewidth,height=2.1in}
\quad\epsfig{file=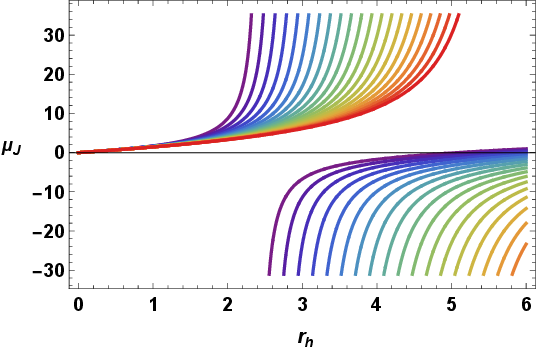,width=.4\linewidth,height=2.2in}
\quad\epsfig{file=m.eps,width=.06\linewidth,height=2.1in}
\caption{\label{jte2}  Behavior of JT coefficient by varying $M$ with $\beta=0.1$ (left plot) and $\beta=0.5$ (right plot) with $\alpha=1$}
\end{figure}

The JT effect is a key thermodynamic phenomenon in which the temperature of a system varies with pressure while maintaining constant enthalpy. Within the framework of extended phase space, the JT expansion of a BH corresponds to an isenthalpic transformation. The JT coefficient ($\mu_J$) characterizes the rate at which the temperature responds to changes in pressure, thereby indicating whether the BH undergoes heating or cooling during the process. Also, by employing the expressions for thermodynamic pressure (\ref{2-7}), entropy (\ref{2-8}), and temperature (\ref{2-12}), one can derive the corresponding EoS for the BH as 
\begin{eqnarray}\label{3-1}
\mu_J={\Big(\frac{\partial T}{\partial P}\Big)}_{H}.
\end{eqnarray}
For the BH system under study, the quantity $H$ is identified as its enthalpy. Also, the following formulas can also be included to determine the BH's EoS: thermodynamic pressure (\ref{2-7}), temperature (\ref{2-12}) and entropy (\ref{2-8}).
\begin{equation}\label{3-2}
\mu_J={\Big(\frac{\partial T}{\partial
P}\Big)}_{M}=\frac{1}{C_P}\Big[T\Big({\frac{\partial V}{\partial
T}}\Big)_P-V\Big].
\end{equation}
The specific heat at fixed pressure, represented as $C_P$, is considered in this context. By imposing the condition $\mu_J = 0$, one directly arrives at the expression for the inversion temperature.
\begin{equation}\label{3-3}
T_{i}=V {\Big(\frac{\partial T}{\partial V}\Big)}_{P}.
\end{equation}

Equations (\ref{2-7}), (\ref{2-8}) and (\ref{2-9}) can be used to express the pressure $P$ as a function of the mass, coupling constant and event horizon radius of the BH, i.e.,
\begin{equation}\label{3-2-1}
P({M},{r}_{h})=-\frac{-6 M+\frac{\alpha  \big(\beta ^2+3 r_{h}^2+3
\beta  r_{h}\big)}{(\beta +r_{h})^3}+3 r_{h}}{8 \pi  r_{h}^3}.
\end{equation}
Placing $P(M,r_h)$ into the BH temperature form shown in Equation (\ref{2-12}) yields
\begin{equation}\label{3-2-2}
T({M},{r}_{h})=\frac{\frac{6 M}{r_{h}}-\frac{\alpha  \big(\beta ^2+3
r_{h}^2+3 \beta  r_{h}\big)}{r_{h} (\beta +r_{h})^3}-\frac{\alpha
r_{h}^2}{(\beta +r_{h})^4}-2}{4 \pi  r_{h}}.
\end{equation}
The JT coefficient $\mu_J$ is determined using the temperature and pressure formulations of the deformed Schwarzschild-AdS BH. These are expressed in terms of the deformation parameter, mass and event horizon as shown by:
\begin{equation}\label{3-2-3}
\mu_J={\Big(\frac{\partial T}{\partial
P}\Big)}_{M}={\Big(\frac{\partial T}{\partial
{r_{h}}}\Big)}_{M}{\Big(\frac{\partial {r_{h}}}{\partial
P}\Big)}_{M}=\frac{\Big(\partial T/\partial
{r_{h}\Big)}_{M}}{\Big(\partial P/\partial {r_{h}}\Big)_{M}}.
\end{equation}
Thus, we get
\begin{eqnarray}\nonumber
\mu_J&=&\bigg[4 r_{h} \big(-6 M (\beta +r_{h})^5+r_{h}^6+6 \alpha
r_{h}^4+\beta ^4 \big(\alpha +5 r_{h}^2\big)+5 \beta ^3 r_{h}
\big(\alpha +2 r_{h}^2\big)+10 \beta ^2 r_{h}^2 \big(\alpha
+r_{h}^2\big)+5 \beta  r_{h}^3 \big(2
\alpha\\\nonumber&+&r_{h}^2\big)+\beta ^5 r_{h}\big)\bigg]\bigg[3
(\beta +r_{h}) \big(\alpha \beta ^3-6 M (\beta +r_{h})^4+2 r_{h}^5+8
\beta r_{h}^4+4 r_{h}^3 \big(\alpha +3 \beta ^2\big)+r_{h}^2 \big(6
\alpha \beta +8 \beta ^3\big)+2 r_{h}\\\label{3-2-3s}&\times& \big(2
\alpha \beta ^2+\beta ^4\big)\big)\bigg]^{-1}.
\end{eqnarray}

Figures~(\ref{jte1}) and~(\ref{jte2}) illustrate the variation of the JT coefficient with the horizon radius for different values of the model parameters. The results demonstrate that the JT response is sensitive to the BH size and the deformation parameters $\alpha$ and $\beta$. In particular, a positive JT coefficient ($\mu_J>0$) indicates cooling during expansion, whereas a negative value ($\mu_J<0$) corresponds to heating. For smaller BHs, the coefficient remains positive, while increasing the horizon radius leads to a transition toward negative values. Furthermore, increasing $\alpha$ and $\beta$ reduces the magnitude of the JT coefficient, indicating that these parameters significantly modify the strength of the JT effect. Overall, the observed behavior highlights the important role of the deformation parameters in governing the thermodynamic response of BHs across different horizon-radius regimes.

\begin{figure}\center
\epsfig{file=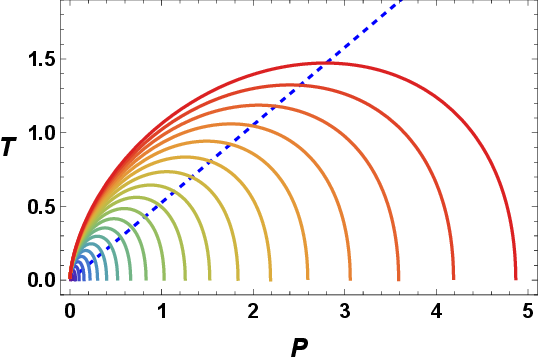,width=.4\linewidth,height=2.2in}
\quad\epsfig{file=m.eps,width=.06\linewidth,height=2.1in}
\quad\epsfig{file=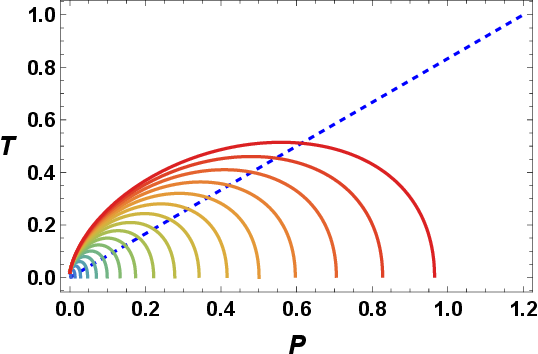,width=.4\linewidth,height=2.2in}
\quad\epsfig{file=m.eps,width=.06\linewidth,height=2.1in}
\caption{\label{h1} Behavior of inversion and isenthalpic curves with varying values of $M$ with $\alpha=1$ (left plot) and $\alpha=1.5$ (right plot) with $\beta=0.02$. The inversion curves are indicated by blue dashed lines and the remaining curves represent the isenthalpic curves.}
\epsfig{file=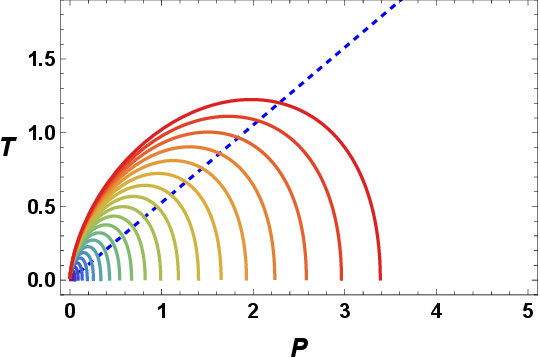,width=.4\linewidth,height=2.2in}
\quad\epsfig{file=m.eps,width=.06\linewidth,height=2.1in}
\quad\epsfig{file=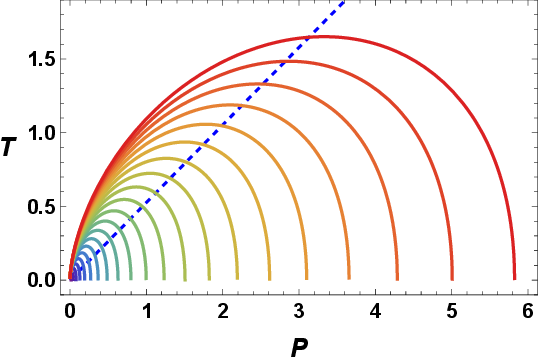,width=.4\linewidth,height=2.2in}
\quad\epsfig{file=m.eps,width=.06\linewidth,height=2.1in}
\caption{\label{h2}  Behavior of inversion and isenthalpic curves with varying values of $M$ with $\beta=0.02$ (left plot) and $\beta=0.04$ (right plot) with $\alpha=1$.}
\end{figure}

By considering $\mu_J=0$ with $P_i=0$ in Eq.(\ref{3-2-3s}), we obtain
\begin{equation}\label{3-2-4}
2 \beta ^5+2 r_{ch}^5+10 \beta  r_{ch}^4+r_{ch}^3 \big(20 \beta ^2-3
\alpha \big)+r_{ch}^2 \big(20 \beta ^3-\alpha  \beta \big)+10 \beta
^4 r_{ch}=0.
\end{equation}
Furthermore, the recognized equation for the inversion temperature of the BH is as follows:
\begin{equation}\label{3-2-4ba}
T_i=\frac{2 P_i r_{ch}}{3}+\frac{\pi ^{3/2} \alpha  r^2_{ch}}{3 \left(\sqrt{\pi } \beta +\sqrt{\pi } r_{ch}\right)^5}-\frac{\pi  \alpha  r_{ch}}{12 \left(\sqrt{\pi } \beta +\sqrt{\pi } r_{ch}\right)^4}-\frac{1}{12 \pi  r_{ch}}.
\end{equation}
When there is no inversion pressure, the minimum inversion temperature is as
\begin{equation}\label{3-2-4ba}
T_i^{min}=\frac{\pi ^{3/2} \alpha  r^2_{ch}}{3 \left(\sqrt{\pi } \beta +\sqrt{\pi } r_{ch}\right)^5}-\frac{\pi  \alpha  r_{ch}}{12 \left(\sqrt{\pi } \beta +\sqrt{\pi } r_{ch}\right)^4}-\frac{1}{12 \pi  r_{ch}}.
\end{equation}

To comprehend the JT effect in BHs, isenthalpic curves must be analyzed using the proper physical parameters. Inversion curves, which connect mass, temperature and entropy, are commonly used in BH investigations to illustrate state-changing circumstances. However, Isenthalpic curves concentrate on energy conservation in phenomena such as interactions between particles or radiation close to the event horizon. Although inversion curves draw attention to phase transitions, thermodynamics is the foundation of both ideas.
For different BH masses with bigger $r_h$ roots, the isenthalpic and inversion curves in the $T$-$P$ plane are shown in Figs. (\ref{h1})–(\ref{h2}). Isenthalpic curves above the inversion curve have positive slopes, separating the cooling and heating zones. The gradient changes during the JT expansion below the inversion curve and when heating takes place, the pressure drops.

During the JT effect, the constant mass curves slope upward due to cooling at lower temperatures and upward due to heating at higher temperatures, as indicated by the curves' negative slope. The heating phase (heat addition) is represented by the region below the inversion curve, while the cooling phase (heat loss) is represented by the region above. Isenthalpic curves have positive slopes in the cooling zone and negative slopes in the heating zone. These graphs show invariant enthalpy, whose derivative with respect to pressure is positive during the cooling phase and negative during the heating phase.
The cooling and heating areas are divided by the inversion curve, which also prevents transitions over it. Temperature changes related to heating or cooling only occur within the associated zones and not at the boundary. Analyzing thermodynamic behavior requires an understanding of the connection between temperature and pressure. The isenthalpic curves are strongly influenced by the deformation parameters. The curves peak for greater BH masses and higher values of $\alpha$, as Fig. (\ref{h1}) illustrates. In contrast, Fig. (\ref{h2}) shows that greater isenthalpic curve peaks are produced by smaller values of the deformation parameter $\beta$.

\section{Conclusion}\label{S5}

In this study, we have conducted a comprehensive test of the \textsl{GBF} associated with the JT expansion of deformed Schwarzschild-AdS BHs. In this case, our results show the behavior of massless scalar fields (KG equation) in the vicinity of these BHs, particularly in relation to the effective potential influenced by various physical parameters, including deformation parameters, mass and angular momentum. Also, we derived the exact effective potential from the transformed KG equation and observed that the deformation parameter enhances the effective potential of Schwarzschild BHs. This enhancement leads to notable variations in the absorption and scattering processes of scalar fields, illustrating the critical role of the effective potential in determining wave behavior near the event horizon. In this case, the relationship between wave frequency and the effective potential is shown by our results, which show that higher-frequency waves are more likely to pass through obstacles while lower-frequency waves are more likely to be reflected. Also, we examined the JT expansion of BH models, showing how temperature ($T$) and pressure affect their thermodynamic properties. Another important result in this BH measurement, showing and illustrating the energy dynamics related to BH radiation and particle behavior near the event horizon, was illustrated by the examination of isenthalpic curves. In this result, using the inversion curve, which illustrates the circumstances in which these processes take place, we succeeded in identifying discrete heating and cooling zones. This study advances our understanding of BH thermodynamics as well as scalar fields (KG) interacting with curved spacetime. Also, our results have ramifications for QFT in curved spacetime and provide a foundation for future research on information loss and energy emission rates in BH physics. {Overall, the \textsl{GBF} and JT analyses provide complementary perspectives on the effects of spacetime deformation. While the \textsl{GBF} reveals how the deformation parameters modify the effective potential and the propagation of scalar waves, the JT analysis demonstrates their influence on the BH's thermodynamic response through isenthalpic and inversion curves. Their combined investigation provides a more comprehensive understanding of the deformed Schwarzschild--AdS BH by examining both its wave-scattering characteristics and its extended thermodynamic behavior.}

\section*{Acknowledgments} 
This work was supported and funded by the Deanship of Scientific Research at Imam Mohammad Ibn Saud Islamic University (IMSIU) (grant number IMSIU DDRSP2502).

\end{document}